\documentclass[aps,prb,reprint,superscriptaddress,longbibliography]{revtex4-1}

\usepackage{graphicx} 
\usepackage{dcolumn}
\usepackage{amsfonts}
\usepackage{amssymb}
\usepackage{amsmath}
\usepackage{bm}
\usepackage{epsf}
\usepackage{epstopdf}
\newcommand{\be}{\begin{equation}}
\newcommand{\ee}{\end{equation}}
\newcommand{\bea}{\begin{eqnarray}}
\newcommand{\eea}{\end{eqnarray}}

\begin{document} 

\title{Quantum degenerate dipolar Fermi gas}

\author{Mingwu Lu}
\affiliation{Department of Physics, University of Illinois at Urbana-Champaign, Urbana, IL 61801}
\affiliation{Department of Applied Physics, Stanford University, Stanford CA 94305}
\affiliation{E. L. Ginzton Laboratory, Stanford University, Stanford CA 94305}

\author{Nathaniel Q. Burdick}
\affiliation{Department of Physics, University of Illinois at Urbana-Champaign, Urbana, IL 61801}
\affiliation{Department of Applied Physics, Stanford University, Stanford CA 94305}
\affiliation{E. L. Ginzton Laboratory, Stanford University, Stanford CA 94305}

\author{Benjamin L. Lev}
\affiliation{Department of Applied Physics, Stanford University, Stanford CA 94305}
\affiliation{E. L. Ginzton Laboratory, Stanford University, Stanford CA 94305}
\affiliation{Department of Physics, Stanford University, Stanford CA 94305}

\date{\today}
\maketitle

{\bf{The interplay between crystallinity and superfluidity is of great fundamental and technological interest in condensed matter settings.  In particular, electronic quantum liquid crystallinity arises in the non-Fermi liquid, pseudogap regime neighboring a cuprate's unconventional superconducting phase~\cite{Fradkin2009}.  While the techniques of ultracold atomic physics and quantum optics have enabled explorations of the strongly correlated, many-body physics inherent in, e.g., the Hubbard model~\cite{bloch:review}, lacking has been the ability to create a quantum degenerate Fermi gas with interparticle interactions---such as the strong dipole-dipole interaction~\cite{PfauReview09}---capable of inducing analogs to electronic quantum liquid crystals.  We report the first  quantum degenerate dipolar Fermi gas, the realization of which opens a new frontier for exploring strongly correlated physics and, in particular, the quantum melting of smectics in the pristine environment provided by the ultracold atomic physics setting~\cite{Kivelson03}.
A quantum degenerate Fermi gas of the most magnetic atom $^{161}$Dy is produced by laser cooling to 10~$\mu$K before sympathetically cooling with ultracold, bosonic $^{162}$Dy.  The temperature of the spin-polarized $^{161}$Dy is a factor $T/T_{F}=0.2$ below the Fermi temperature $T_{F}=300$~nK.  The co-trapped $^{162}$Dy concomitantly cools to approximately $T_{c}$ for Bose-Einstein condensation, thus realizing a novel, nearly quantum degenerate dipolar Bose-Fermi gas mixture.  }}

Quantum soft  phases are states of quantum matter intermediate between canonical states of order and disorder, and may be considered the counterparts of liquid crystalline and glassy states in classical (soft) condensed matter physics.  Such phases tend to arise under competition between short and long-range interactions and often result in the non-Fermi liquid, strongly correlated behavior manifest in some of the most interesting electronic materials of late:  high-\emph{T}$_c$ cuprate superconductors, strontium ruthenates, 2D electron gases, and iron-based superconductors~\cite{Fradkin2010}.   Recent theory suggests the long-range, anisotropic dipole-dipole interaction (DDI) among atoms in a degenerate Fermi gas, confined in an harmonic trap or optical lattice, may also induce transitions to states beyond the now-familiar insulating, metallic, and superfluid.   Namely, phases that break rotational, translational, or point group symmetries may emerge in a manner akin to those found in classical liquid crystals, e.g., the nematic and smectic~\cite{Fradkin2009}.  

\begin{figure*}[t]
\begin{center}
 \includegraphics[width=0.975\textwidth]{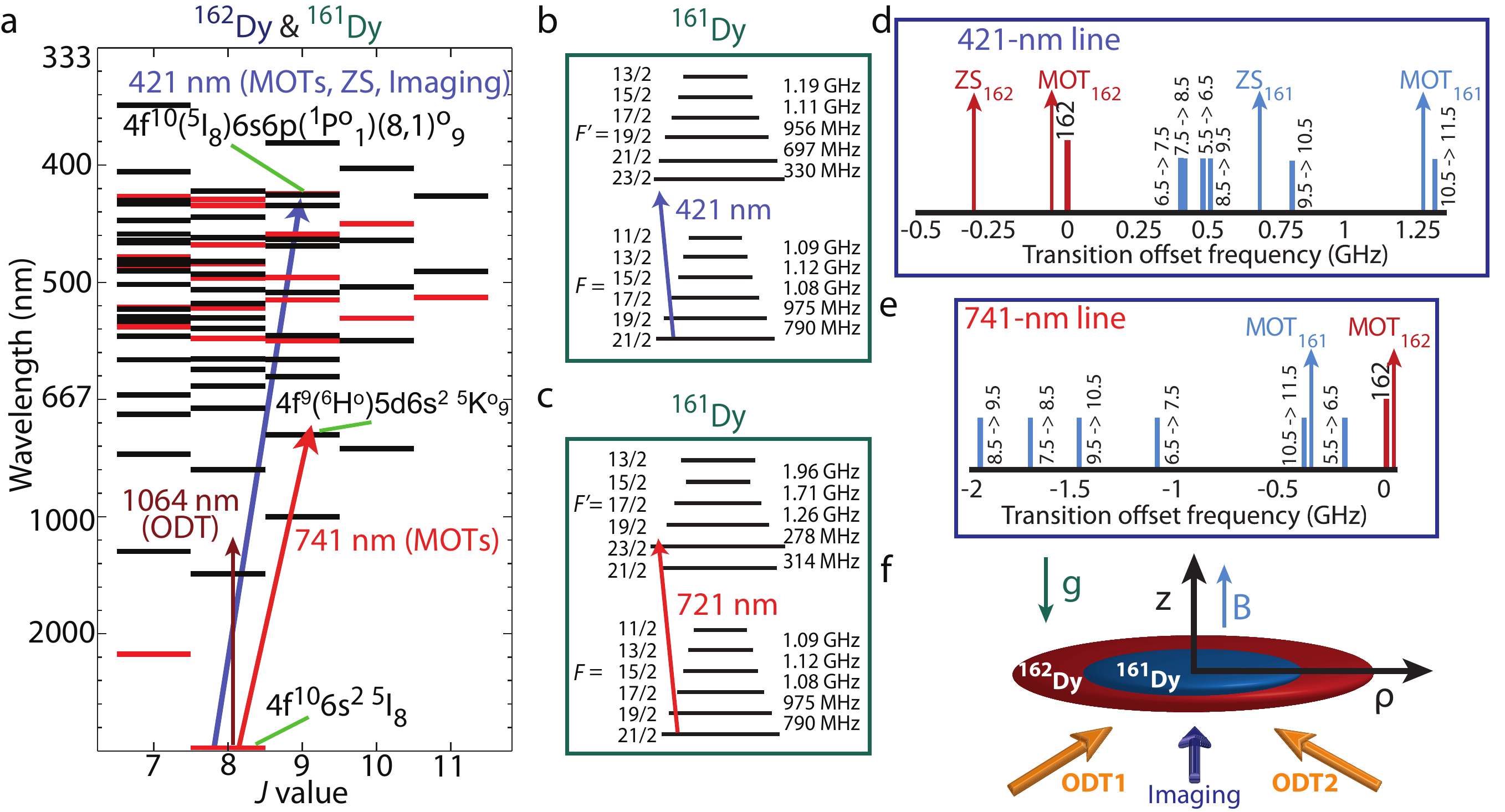}

\caption{\label{States} {\textbf{Dy transitions and laser cooling and trapping scheme}}. {\textbf{a}}, Electronic energy level structure for bosonic $^{162}$Dy (nuclear spin $I=0$) and fermionic  $^{161}$Dy ($I=5/2$), including laser cooling and trapping transitions.  {\textbf{b}}, Additional ground and excited-state hyperfine structure exists for $^{161}$Dy  ($F = I + J$, where $J=8$ is the total electronic angular momentum and primes denote the excited states). Shown is the 32-MHz-wide transition at 412 nm used for the transverse cooling, Zeeman slower, capture MOT, and imaging beams.  {\textbf{c}}, Blue-detuned, narrow-line (1.8 kHz-wide) MOT cooling transition at 741 nm.  {\textbf{d}}, Spectra of 421-nm $^{162}$Dy and  $^{161}$Dy (hyperfine) transitions including relative detunings of each MOT and Zeeman slower (ZS) laser. {\textbf{e}}, Transition and detuning spectra for MOT on 741-nm line. {\textbf{f}}, Sketch of dual species crossed optical dipole trap (ODT) aspect ratio along with magnetic field $B$ and  gravity $g$ orientation. The imaging (421-nm) beam and the orthogonal ODT (1064-nm) beams are in the $\hat{\rho}$-plane.}
\end{center}
\end{figure*}

Degenerate gases of highly magnetic fermionic atoms, such as $^{161}$Dy, may shed light on QLC physics without unwanted solid state material complexity, disorder, and dynamical lattice distortions.  Uniaxial (meta-nematic)~\cite{Miyakawa:2008} and biaxial nematic~\cite{Fregoso:2009} distortions of the Fermi surface of a harmonically trapped gas in the presence of a polarizing field may be observable as well as meta-nematic and smectic phases in 2D anisotropic optical lattices~\cite{Quintanilla:2009,Liu10,DasSarma10c}.  An exciting prospect lies in the possibility of achieving spontaneous magnetization in dipolar systems coupled with nematic order~\cite{Fregoso2009b,fregoso:2010}.  Additionally, DDI-induced pairing of fermions may lead to supersolidity~\cite{Hofstetter11} and bond order solids~\cite{Clark11}.  

However, obtaining a quantum degenerate dipolar Fermi gas has been a difficult, unrealized experimental challenge.  The highly magnetic fermionic atoms $^{53}$Cr (6 Bohr magnetons $\mu_{B}$) and $^{167}$Er (7~$\mu_{B}$) have yet to be cooled below 10 $\mu$K~\cite{Gorceix06,Berglund:2007}.   The fermionic polar molecule $^{40}$K$^{87}$Rb (0.57~Debye) has been cooled to near degeneracy ($T/T_{F}=1.4$)~\cite{Ni2010}  and loaded into a long-lived lattice while partially polarized (0.2~D)~\cite{Chotia11}, but complexities arising from ultracold chemistry have hampered additional evaporative cooling~\cite{Ni2010}.  
In contrast, magnetic fermionic atoms do not undergo chemical reactions and are immune to inelastic dipolar collisions when spin polarized in high magnetic fields~\cite{Lu2011b,Pasquiou10}.  

The strong, $r^{-3}$ character of the DDI arises in ground state polar molecules though a polarizing electric field that mixes opposite parity states.  This electric field breaks rotational symmetry; consequently, observing the full range of true (non-meta) quantum nematic and fluctuating smectic phases, and their often unusual topological defects, is not possible in systems of fermionic polar molecules, especially in three dimensions.  By contrast, highly magnetic atoms exhibit the DDI interaction even in the absence of a polarizing field.  Moreover, the magnetic DDI can be tuned from positive to negative~\cite{Pfau02}, which may be important for simulating dense nuclear matter.  

Dysprosium's isotopes $^{161}$Dy and $^{163}$Dy are the most magnetic fermionic atoms.  With a dipole moment of $\mu = 10$~$\mu_{B}$, $^{161}$Dy provides a DDI length $l_{\text{DDI}} = \mu_{0}\mu^{2}m/4\pi\hbar^{2}$ that is factors of $[400,8,2]$ larger than that of [$^{40}$K, $^{53}$Cr, $^{167}$Er].  With respect to fully saturated $^{40}$K$^{87}$Rb (0.57 D), $^{161}$Dy is 30$\times$ less dipolar for equal densities, but within a factor of 2 if confined in a lattice of less than half the periodicity.  Lattices of wavelength 400--500~nm may be possible with Dy, whereas for molecules, photon scattering from rovibronic states at these wavelengths may reduce gas lifetimes.  Indeed, $^{161}$Dy confined in a 450-nm lattice would provide a DDI strength more than 3$\times$ larger than $^{40}$K$^{87}$Rb confined in a 1064-nm lattice, if the electric dipole moment is unsaturated (0.2 D) to maintain collisional stability~\cite{Chotia11}.  We estimate that the DDI strength of Dy confined in short-wavelength lattices would be sufficient to observe some of the exotic many-body physics recently proposed~\cite{Pupillo11,Hofstetter11,Clark11,Gorshkov11}.

Until recently, the laser cooling of Dy posed an insurmountable challenge due to its complex internal structure and the limited practicability of building repumping lasers: $\sim$140 metastable states exist between the ground state and the broadest laser cooling transition at 421 nm (see Fig.~\ref{States}a).  Moreover, an open $f$-shell submerged underneath closed $s$-shells, combined with a large magnetic moment and electrostatic anisotropy from the $L=6$ orbital angular momentum, pose challenges to molecular coupled-channel calculations~\cite{Svetlana11} which could otherwise guide early experiments.  Even more daunting is the additional hyperfine structure of the fermions, shown in Fig.~\ref{States}b--c, which splits each level into six.   

Despite this vast energy level state-space, a repumperless magneto-optical (MOT) technique  was able to individually laser cool and trap of all five high-abundance isotopes (three bosons, two fermions) to 1~mK using a single laser~\cite{Lu2010} (see Methods for details).  Moreover, the complex homonuclear molecular potentials---involving 153 Born-Oppenheimer surfaces---and the associated multitude of scattering lengths did not inhibit the efficient Bose-condensation of spin-polarized $^{164}$Dy through forced evaporative cooling~\cite{Lu2011b}. 

However, quantum degeneracy of identical Fermi gases is often more difficult to achieve than Bose-condensation because $s$-wave collisions are forbidden due to the requirement that the total wave function for two identical fermions be anti-symmetric with respect to particle exchange.  Rethermalization from elastic collisions cease below the threshold for $p$-wave collisions (at typically 10--100 $\mu$K), and efficient evaporative cooling can no longer be maintained.  Co-trapping mixtures of particles---either as different spin states of the same atom or as mixtures of isotopes or elements---reintroduces $s$-wave collisions, providing a finite elastic cross-section for scattering even down to low temperatures.  The mixture needs to be stable against inelastic collisions which could add heat or induce trap loss.  Evaporating $^{161}$Dy in a mixed state of two spins, as proved efficient for $^{40}$K,~\cite{DeMarcoThesis} would lead to large dipolar relaxation-induced heating even in the presence of  small, mG-level fields because the inelastic, single spin-flip cross-section $\sigma_{1} = \sigma\zeta(k_{f}/k_{i})$ scales strongly with dipole moment~\cite{Hensler:2003}: \be \sigma =  \frac{8\pi}{15}F_{1}F_{2}^{2} \left(\frac{\mu_{0}g_{1}g_{2}\mu_{B}^{2}m}{4\pi\hbar^{2}}\right)^{2}\frac{k_{f}}{k_{i}}, \nonumber\ee where $F_{1}$ is the spin of the atom whose spin flips ($F_{1}=F_{2}$ for identical particles), $g_{i}$ are $g$-factors for atom $i$,  $m$ is mass, and $k_{i}$ and $k_{f}$ are the initial and final momenta.  For $^{161}$Dy ($^{162}$Dy), $F=21/2$ ($F = J = 8$) and $g_{F}$ ($g_{J}$) = 0.95 (1.24). The function $\zeta(k_{f}/k_{i}) = [1 +\epsilon h(k_{f}/k_{i})]$, where $\epsilon = \pm1,0$ and $h(x)$ is defined in Methods, accounts for the contributions of even, odd, or all partial waves to the scattering process.  
\begin{figure}[t]
\begin{center}
\includegraphics[width=0.475\textwidth]{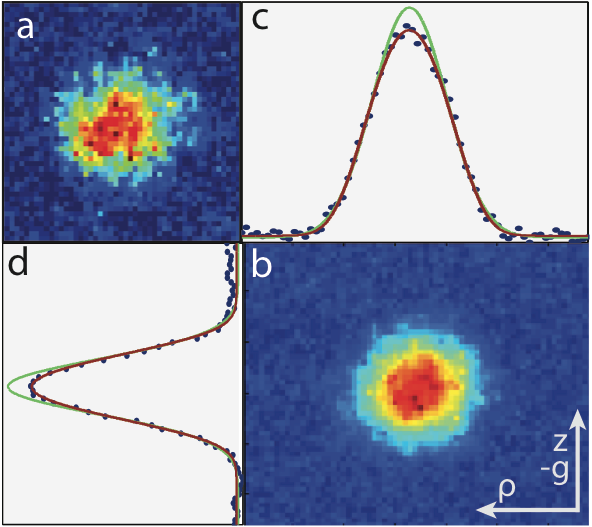}
\caption{\label{TOF} {\textbf{Images of the Dy degenerate Fermi gas}}.  {\textbf{a}},  Single shot time-of-flight absorption image at  $t=6$~ms. {\textbf{b}}, Average of six  images. Density integrations versus $\hat{\rho}$ ({\textbf{c}}) and $\hat{z}$ ({\textbf{d}}).  The green curve is a gaussian fit to the data's wings (radius $\sigma=20$ $\mu$m), while the red curve is a fit to a Thomas-Fermi distribution.  Data are consistent with a Thomas-Fermi distribution of  $T/T_{F}=0.21(5)$.  The Fermi velocity and temperature are 5.6(2) mm/s and $306(20)$ nK, respectively, and the gas temperature is $64(16)$ nK.  The degenerate Fermi gas contains $6.0(6)$$\times$$10^{3}$ atoms at peak density $4(1)\times 10^{13}~\text{cm}^{-3}$. 
}
\end{center}
\end{figure} 

We choose, therefore, to seek a degenerate dipolar Fermi gas with Dy by sympathetically cooling $^{161}$Dy with the boson $^{162}$Dy while both are spin-polarized in their strong-magnetic-field seeking ground states:  $|F,m_{F}\rangle=|21/2,-21/2\rangle$ for $^{161}$Dy and $|J,m_{J}\rangle=|8,-8\rangle$ for $^{162}$Dy. See Fig.~\ref{States}a--c for energy level schemes. Preparation of this ultracold Bose-Fermi mixture---the first such mixture for strongly dipolar species---builds on our single-species technique~\cite{Lu2011b} for Bose-condensing $^{164}$Dy and relies on the laser cooling and trapping of two isotopes before loading both into an optical dipole trap (ODT) for forced evaporative cooling.    We sketch here the experimental procedure; further details are provided in the Methods.

Isotopes $^{161}$Dy and $^{162}$Dy are collected sequentially in a repumperless MOT operating on the 421-nm transition~\cite{Lu2010} (Fig.~\ref{States}d), with final MOT populations of $N = 2\times 10^7$ and $4\times 10^7$, respectively.  Next, simultaneous narrow-line, blue-detuned MOTs~\cite{Berglund:2008,Lu2011b} cool both isotopes to 10~$\mu$K via the 741-nm transition (Fig.~\ref{States}e) for 5 s to allow any remaining metastable atoms to decay to the ground state. The blue-detuned MOTs also serve to spin polarize~\cite{Berglund:2008,Lu2011b} both isotopes to their maximally high-field-seeking (metastable) states $m_{F}=+F$ ($m_{J}=+J$) for $^{161}$Dy ($^{162}$Dy).

\begin{figure*}[t]
\begin{center}
 \includegraphics[width=0.975\textwidth]{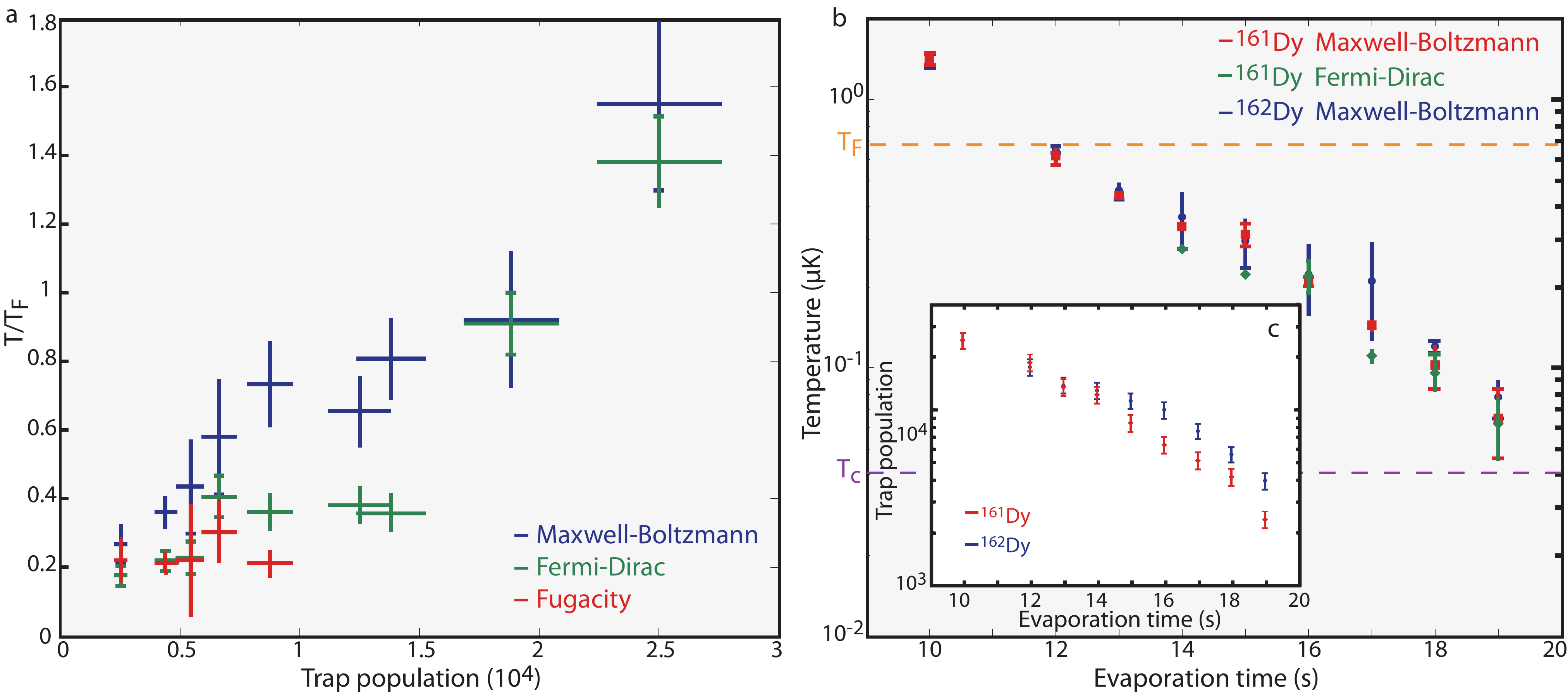}
\caption{\label{TTF}  {\textbf{Temperature and population of $^{161}$Dy and $^{162}$Dy mixture.}} {\textbf{a}}, Three measures (see text) of $^{161}$Dy $\hat{\rho}$ temperature--to--Fermi temperature  $T_{F}$  as trap population decreases due to forced evaporation of the spin-polarized Bose-Fermi mixture.  {\textbf{b}}, The dipolar Bose-Fermi mixture remains in thermal contact throughout the evaporation sequence as measured by fits of TOF densities to gaussian or TF distributions.  The orange dashed line demarcates the boundary below which the temperature of $^{161}$Dy is lower than $T_{F}$.  Likewise, the purple dashed line demarcates the temperature below which $T_{c}$ for $^{162}$Dy Bose degeneracy would be reached given  the trap   frequencies and population at 19 s. {\textbf{c}}, Trap populations of the  spin-polarized Bose-Fermi mixture versus evaporation time.  
}
\end{center}
\end{figure*} 

The blue-detuned MOTs of the two isotopes can be spatially separated due to the dependence of the MOTs' positions on laser detuning~\cite{Berglund:2008,Lu2011b}. This allows the isotopes to be sequentially loaded into the 1064-nm ODT1 in Fig.~\ref{States}f, which is aligned above the $^{161}$Dy MOT but below the $^{162}$Dy MOT.  First $^{162}$Dy and then $^{161}$Dy is loaded into ODT1 by shifting the quadrupole center with a vertical bias field.  All 741-nm light is extinguished before the spin of both isotopes are rotated via radiofrequency (RF) adiabatic rapid passage (ARP) into their absolute ground states $m_{F}=-F$ ($m_{J}=-J$) for $^{161}$Dy ($^{162}$Dy).      The ODT1 populations of $^{161}$Dy and $^{162}$Dy are both initially $1\times10^{6}$ before plain evaporation cools the gases to 1--2 $\mu$K within 1~s.   A 0.9 G field is applied close to the trap axis of symmetry $\hat{z}$ throughout plain and forced evaporation.  This provides a $\Delta m = 1$ Zeeman shift equivalent to $50$ (70)~$\mu$K for $^{161}$Dy ($^{162}$Dy).  Because this is much larger than the temperatures of the gases, the field serves to maintain spin polarization while stabilizing the strongly dipolar $^{162}$Dy Bose gas against collapse as its phase-space density increases~\cite{Lu2011b}.  

Magnetic Stern-Gerlach measurements and observations of fluorescence versus polarization are consistent with an RF ARP sequence that achieves a high degree of spin purity for each isotope.   Remnant population in metastable Zeeman substates quickly decays to the absolute ground state via  dipolar relaxation  regardless of collision partner at a rate of $\Gamma \propto \sigma_{1}n\bar{v} = 1\text{--}10$~s$^{-1}$, where $n$ is the atomic density and $\bar{v}$ is the relative velocity during the plain evaporation stage.   (Since $g_{F}F = g_{J}J$, collisions between Bose-Bose, Bose-Fermi, and Fermi-Fermi pairs result in $\Gamma$'s of similar magnitude as long as $k_{i}\approx k_{f}$ since $h(x$$\rightarrow$$1) $$\rightarrow$$ 0$.  This condition is fulfilled during plain evaporation due to a low ratio of Zeeman--to--kinetic energy.  For example,  inelastic dipolar $^{161}$Dy--$^{161}$Dy collisions ($\epsilon = -1$) proceed at rate $\Gamma = 1$--5~s$^{-1}$ even in the absence of $^{162}$Dy.)  Thus, a (two-body) collisionally stable mixture of identical bosons and identical fermions is prepared within the 1 s between spin rotation and forced evaporation.

Subsequently crossing ODT1 with ODT2 forms an oblate trap with frequencies $[f_{x},f_{y},f_{z}]= [500, 580, 1800]$~Hz.  Ramping down the optical power lowers the trap depth and evaporates spin-polarized $^{161}$Dy to quantum degeneracy; with a 19 s evaporation, the final trap has frequencies [180, 200, 720]~Hz and $\bar{\omega}/2\pi$ is defined as their geometric mean. Figure~\ref{TOF} shows the density profile of ultracold  $^{161}$Dy, which is more consistent with a Thomas-Fermi (TF) distribution arising from Fermi-Dirac statistics than a gaussian arising from a classical, Maxwell-Boltzmann (MB) distribution.  

Figure~\ref{TTF} shows a collection of such fits as a function of trap population and evaporation time.  The $T/T_{F}$ data are extracted from density profiles using three methods:  1) The high momentum wings are fit to a MB distribution with $k_{B}T_{F}=\hbar\bar{\omega}(6N)^{1/3}$  extracted from measured trap parameters and population (Fig.~\ref{TTF}c); 2) similar to (1), but using a TF distribution; 3) fitting the fugacity to directly extract $T/T_{F}$.  The last method is known to be inaccurate at higher temperatures, while the gaussian fit tends to overestimate the temperature below $T/T_{F}=1$.~\cite{DeMarcoThesis}   The evaporation does not yet seem to reach a plateau in cooling efficiency at 19~s; poor imaging signal-to-noise hampers measurements at longer evaporation times. 

Data in Fig.~\ref{TTF}b show that thermal equilibrium between the bosons and fermions is maintained throughout the evaporation, and Bose-condensation of $^{162}$Dy within the mixture is nearly reached for an evaporation of 19~s.  We estimate the corresponding critical temperature $T_{c}\approx40$ nK of co-trapped $^{162}$Dy by scaling the measured $T_{c}\approx120$ nK  of singly trapped $^{162}$Dy with the cube root of their relative trap populations.  ($^{162}$Dy has been Bose-condensed in the absence of $^{161}$Dy; manuscript in preparation.)  The nearly doubly degenerate dipolar Bose-Fermi mixture may lead to interesting dipolar and many-body physics once cooling efficiency improves.  

While interacting BECs invert their anisotropic aspect ratio upon time-of-flight expansion, anisotropic degenerate Fermi gases tend to a spherical shape.  As the DDI strength increases, the degenerate Fermi gas will expand into a prolate ellipsoid oriented along the magnetization direction regardless of trap aspect ratio.  Furthermore, the gas may become unstable when the quantity $\epsilon_{dd} = \mu_{0}\mu^{2}(m^{3}\bar{\omega}/16\pi^{2}\hbar^{5})^{1/2}>1$.~\cite{Pu08}  At the lowest attained $T/T_{F}$, $\epsilon_{dd} = 0.2$ for $^{161}$Dy, and the ratio is $l_{DDI}/l_{F}$ is 0.05. This DDI strength should lead to Fermi surface distortions (as yet unmeasured) at the percent level~\cite{Pu08}.  Both ratios could be enhanced  $\sim$3$\times$ by increasing trap frequency using a more spherical confinement---while maintaining stability of the dipolar Bose gas---and by increasing trap population.  Additionally, the efficiency of evaporation---from gaussian fits to the  data, $\gamma = 3\frac{d(\ln{T/T_{F}})}{d(\ln{N})} \approx 2.3$---may be improved by optimally tuning the magnetic bias field near one of the presumably large number of Feshbach resonances.   

Surprisingly, we achieve the forced evaporative cooling of spin-polarized $^{161}$Dy {\textit{without}}  $^{162}$Dy  to $T/T_{F}=0.7$ at $T_{F}=500$~nK.  As mentioned above, achieving quantum degeneracy with spin-polarized identical fermions alone is usually not possible due to suppression of elastic scattering below the $p$-wave threshold 50~$\mu$K.~\cite{Svetlana11}     That such a low temperature ratio is achieved may be a novel consequence of the highly dipolar nature of this gas:  namely, that a significant elastic cross-section persists to low temperatures due to the as yet unobserved phenomenon of universal dipolar scattering~\cite{Bohn09}.  The associated scattering rate is expected to scale as $m^{3/2}\mu^{4}$ regardless of the details of the short-range molecular potential. The predicted fermionic Dy universal dipolar cross-section, $7.2\times10^{-12}$ cm$^{2}$, is nearly equal to $^{87}$Rb's $s$-wave cross-section and could provide sufficient rethermalization for the evaporative cooling we observe.  While future measurements will quantify spin purity, Fermi statistics does not inhibit identical fermions from spin purifying via dipolar relaxation once the RF ARP sequence populates the majority of the atoms in the absolute ground state.  We will present these data along with supporting measurements of collisional cross-sections and scattering lengths in a future manuscript.

The efficacy of experimental proposals~\cite{Kivelson03} for studying the quantum melting of QLCs---important for better understanding the relationship of incipient electronic stripe order to unconventional superconductivity---may now be investigated using degenerate $^{161}$Dy in optical lattices wherein $l_{DDI}/l_{F}$ can be greatly enhanced~\cite{Pupillo11}. Looking beyond QLC physics, the large spin $F=21/2$ of the novel degenerate dipolar Fermi gas presented here opens avenues to explore exotic spinor physics as well as physics associated with strong spin-orbit coupling.

\small{
\vspace{-4mm}
\section{Methods}
\vspace{-2mm}
\noindent \textbf{Repumperless MOT.} The 421-nm Zeeman slower laser is detuned -650 MHz from resonance, and a 421-nm MOT collects $2\times 10^7$ $^{161}$Dy atoms in 4 s with the aid of a pre-slower transverse cooling stage. The MOT operates without repumpers despite many decay channels because the highly magnetic, metastable Dy remains trapped in the magnetic quadruple trap for a sufficiently long time to decay to the ground state~\cite{Lu2010}. All beams derived from the 421-nm laser are then shifted down in frequency by 1.18 GHz and a  $^{162}$Dy MOT collects $4\times 10^7$ atoms in 100 ms. This scheme balances the need for preservation of $^{161}$Dy population against collisional decay~\cite{Lu2010} with the trapping of sufficient $^{162}$Dy for subsequent sympathetic cooling. Reversing this procedure results in fewer atoms due to heating of $^{162}$Dy in the magnetic trap by light scattered from the $^{161}$Dy Zeeman slower beam.
\\

\vspace{-2mm}
\noindent \textbf{Narrow-line MOT}. The narrow-line MOT lasers are blue-detuned 0.6 MHz from the 2-kHz wide 741-nm transition.  The MOTs form below the quadrupole center such that the atomic transition is Zeeman-shifted to the blue of the laser.  The MOT positions are determined by the balance of optical, magnetic, and grativational forces~\cite{Berglund:2008,Lu2011b}, and because the laser detunings for $^{161}$Dy and $^{162}$Dy can be adjusted independently, the two clouds can be easily separated.
\\

\vspace{-2mm}
\noindent \textbf{ODT, RF ARP and crossed ODT}. ODT1 has a waist of 30 (60) $\mu$m and ODT2 has a waist of 20 (60) $\mu$m in $\hat{z}$ $(\hat{\rho})$. ODT1 is ramped on in 100 ms after 5 s of narrow-line cooling. After loading ODT1---120 (800) ms for $^{162}$Dy ($^{161}$Dy)---a 4.3~G field is applied while a 20-ms RF sequence flips the spin of both isotopes to the absolute ground state.  Dipolar relaxation further spin purifies the gas, and within the cross section $\sigma_{1}$, $h(x)=-\frac{1}{2}-\frac{3}{8}\frac{(1-x^{2})^{2}}{x(1+x^{2})}\log\left(\frac{(1-x)^{2}}{(1+x)^{2}}\right)$ [see Ref.~\cite{Hensler:2003}]).  After spin purification, ODT2 is turned on, forming a crossed ODT with depth 300~$\mu$K. ODT1 and ODT2 have initial powers of 18 W and 12 W.  The background limited lifetime exceeds 180 s.  
\\ 

\vspace{-2mm}
\noindent \textbf{Forced evaporation scheme}. The two beams of the crossed ODT are ramped down according to the functional form $P(t) =  P_0 / (1+t/\tau)^\beta$ using experimentally determined parameters $\tau = 6$ s and $\beta = 1.5$ for both beams. Forced evaporation with ODT2 begins 5 s after ODT1. The beam powers are held at the final values for 400 ms after evaporation to allow for equilibration.\\

\vspace{-2mm}
\noindent \textbf{Temperature fitting}. The TOF images are fit to both gaussian and TF profiles, the latter of the form~\cite{DeMarcoThesis}
\begin{equation}
\left[\frac{A}{Li_2(-\zeta)}\right]Li_2\left(-\zeta e^{\frac{-(y-y_0)^2}{2\sigma_{y}^2}}e^{\frac{-(x-x_0)^2}{2\sigma_{x}^2}}\right) + c,
\end{equation}
with fitting parameters $A$, $\zeta$, $y_0$, $x_0$, $\sigma_{y}$, $\sigma_{x}$, and $c$. 
The fitting is scaled by the constant $Li_2(-\zeta)$ such that parameter $A$ corresponds exactly to the image peak OD.  Because the high-velocity component of the cloud is less sensitive to signatures of quantum degeneracy, only the wings of the expanded cloud are used for the gaussian TOF fits.  From the TF fits, the temperature is determined from size of the cloud, $\sigma_{TF,i}^2 = (k_bT/{m\omega_i^2})[1+(\omega_it)^2]$, where $\omega_i$ is the trap frequency in $\hat{z}$ or $\hat{\rho}$  and $t$ is the expansion time. The cloud fugacity $\zeta$ is also determined by the TF fits and provides a direct measure of $T/T_F = [-6Li_3(-\zeta)]^{-1/3}$, where $Li_3$  is the third-order polylogarithm function.

\vspace{-3mm}
\section{Acknowledgments}
\vspace{-3mm}
We thank S.-H.~Youn for early assistance with experiment construction and J.~Bohn, S.~Kotochigova, E.~Fradkin, N.~Goldenfeld, S.~Kivelson, G.~Baym, C. Wu, H. Zhai, X.~Cui, and S. Gopalakrishnan for enlightening discussions.  We acknowledge support from the NSF, AFOSR, ARO-MURI on Quantum Circuits, and the Packard Foundation. }


\begin{thebibliography}{30}%
\makeatletter
\providecommand \@ifxundefined [1]{%
 \@ifx{#1\undefined}
}%
\providecommand \@ifnum [1]{%
 \ifnum #1\expandafter \@firstoftwo
 \else \expandafter \@secondoftwo
 \fi
}%
\providecommand \@ifx [1]{%
 \ifx #1\expandafter \@firstoftwo
 \else \expandafter \@secondoftwo
 \fi
}%
\providecommand \natexlab [1]{#1}%
\providecommand \enquote  [1]{``#1''}%
\providecommand \bibnamefont  [1]{#1}%
\providecommand \bibfnamefont [1]{#1}%
\providecommand \citenamefont [1]{#1}%
\providecommand \href@noop [0]{\@secondoftwo}%
\providecommand \href [0]{\begingroup \@sanitize@url \@href}%
\providecommand \@href[1]{\@@startlink{#1}\@@href}%
\providecommand \@@href[1]{\endgroup#1\@@endlink}%
\providecommand \@sanitize@url [0]{\catcode `\\12\catcode `\$12\catcode
  `\&12\catcode `\#12\catcode `\^12\catcode `\_12\catcode `\%12\relax}%
\providecommand \@@startlink[1]{}%
\providecommand \@@endlink[0]{}%
\providecommand \url  [0]{\begingroup\@sanitize@url \@url }%
\providecommand \@url [1]{\endgroup\@href {#1}{\urlprefix }}%
\providecommand \urlprefix  [0]{URL }%
\providecommand \Eprint [0]{\href }%
\providecommand \doibase [0]{http://dx.doi.org/}%
\providecommand \selectlanguage [0]{\@gobble}%
\providecommand \bibinfo  [0]{\@secondoftwo}%
\providecommand \bibfield  [0]{\@secondoftwo}%
\providecommand \translation [1]{[#1]}%
\providecommand \BibitemOpen [0]{}%
\providecommand \bibitemStop [0]{}%
\providecommand \bibitemNoStop [0]{.\EOS\space}%
\providecommand \EOS [0]{\spacefactor3000\relax}%
\providecommand \BibitemShut  [1]{\csname bibitem#1\endcsname}%
\let\auto@bib@innerbib\@empty
\bibitem [{\citenamefont {Fradkin}\ \emph {et~al.}(2010)\citenamefont
  {Fradkin}, \citenamefont {Kivelson}, \citenamefont {Lawler}, \citenamefont
  {Eisenstein},\ and\ \citenamefont {Mackenzie}}]{Fradkin2009}%
  \BibitemOpen
  \bibfield  {author} {\bibinfo {author} {\bibfnamefont {E.}~\bibnamefont
  {Fradkin}}, \bibinfo {author} {\bibfnamefont {S.~A.}\ \bibnamefont
  {Kivelson}}, \bibinfo {author} {\bibfnamefont {M.~J.}\ \bibnamefont
  {Lawler}}, \bibinfo {author} {\bibfnamefont {J.~P.}\ \bibnamefont
  {Eisenstein}}, \ and\ \bibinfo {author} {\bibfnamefont {A.~P.}\ \bibnamefont
  {Mackenzie}},\ }\bibfield  {title} {\enquote {\bibinfo {title} {Nematic
  {Fermi} fluids in condensed matter physics},}\ }\href@noop {} {\bibfield
  {journal} {\bibinfo  {journal} {Annu. Rev. Condens. Matter Phys.}\ }\textbf
  {\bibinfo {volume} {\textbf{1}}},\ \bibinfo {pages} {153--178} (\bibinfo
  {year} {2010})}\BibitemShut {NoStop}%
\bibitem [{\citenamefont {Bloch}\ \emph {et~al.}(2008)\citenamefont {Bloch},
  \citenamefont {Dalibard},\ and\ \citenamefont {Zwerger}}]{bloch:review}%
  \BibitemOpen
  \bibfield  {author} {\bibinfo {author} {\bibfnamefont {I.}~\bibnamefont
  {Bloch}}, \bibinfo {author} {\bibfnamefont {J.}~\bibnamefont {Dalibard}}, \
  and\ \bibinfo {author} {\bibfnamefont {W.}~\bibnamefont {Zwerger}},\
  }\bibfield  {title} {\enquote {\bibinfo {title} {Many-body physics with
  ultracold gases},}\ }\href {\doibase 10.1103/RevModPhys.80.885} {\bibfield
  {journal} {\bibinfo  {journal} {Rev. Mod. Phys.}\ }\textbf {\bibinfo {volume}
  {80}},\ \bibinfo {pages} {885--964} (\bibinfo {year} {2008})}\BibitemShut
  {NoStop}%
\bibitem [{\citenamefont {Lahaye}\ \emph {et~al.}(2009)\citenamefont {Lahaye},
  \citenamefont {Menotti}, \citenamefont {Santos}, \citenamefont {Lewenstein},\
  and\ \citenamefont {Pfau}}]{PfauReview09}%
  \BibitemOpen
  \bibfield  {author} {\bibinfo {author} {\bibfnamefont {T.}~\bibnamefont
  {Lahaye}}, \bibinfo {author} {\bibfnamefont {C.}~\bibnamefont {Menotti}},
  \bibinfo {author} {\bibfnamefont {L.}~\bibnamefont {Santos}}, \bibinfo
  {author} {\bibfnamefont {M.}~\bibnamefont {Lewenstein}}, \ and\ \bibinfo
  {author} {\bibfnamefont {T.}~\bibnamefont {Pfau}},\ }\bibfield  {title}
  {\enquote {\bibinfo {title} {The physics of dipolar bosonic quantum gases},}\
  }\href@noop {} {\bibfield  {journal} {\bibinfo  {journal} {Rep. Prog. Phys.}\
  }\textbf {\bibinfo {volume} {72}},\ \bibinfo {pages} {126401} (\bibinfo
  {year} {2009})}\BibitemShut {NoStop}%
\bibitem [{\citenamefont {Kivelson}\ \emph {et~al.}(2003)\citenamefont
  {Kivelson}, \citenamefont {Bindloss}, \citenamefont {Fradkin}, \citenamefont
  {Oganesyan}, \citenamefont {Tranquada}, \citenamefont {Kapitulnik},\ and\
  \citenamefont {Howald}}]{Kivelson03}%
  \BibitemOpen
  \bibfield  {author} {\bibinfo {author} {\bibfnamefont {S.~A.}\ \bibnamefont
  {Kivelson}}, \bibinfo {author} {\bibfnamefont {I.~P.}\ \bibnamefont
  {Bindloss}}, \bibinfo {author} {\bibfnamefont {E.}~\bibnamefont {Fradkin}},
  \bibinfo {author} {\bibfnamefont {V.}~\bibnamefont {Oganesyan}}, \bibinfo
  {author} {\bibfnamefont {J.~M.}\ \bibnamefont {Tranquada}}, \bibinfo {author}
  {\bibfnamefont {A.}~\bibnamefont {Kapitulnik}}, \ and\ \bibinfo {author}
  {\bibfnamefont {C.}~\bibnamefont {Howald}},\ }\bibfield  {title} {\enquote
  {\bibinfo {title} {How to detect fluctuating stripes in the high-temperature
  superconductors},}\ }\href {\doibase 10.1103/RevModPhys.75.1201} {\bibfield
  {journal} {\bibinfo  {journal} {Rev. Mod. Phys.}\ }\textbf {\bibinfo {volume}
  {75}},\ \bibinfo {pages} {1201--1241} (\bibinfo {year} {2003})}\BibitemShut
  {NoStop}%
\bibitem [{\citenamefont {Fradkin}\ and\ \citenamefont
  {Kivelson}(2010)}]{Fradkin2010}%
  \BibitemOpen
  \bibfield  {author} {\bibinfo {author} {\bibfnamefont {E.}~\bibnamefont
  {Fradkin}}\ and\ \bibinfo {author} {\bibfnamefont {S.~A.}\ \bibnamefont
  {Kivelson}},\ }\bibfield  {title} {\enquote {\bibinfo {title} {Electron
  nematic phases proliferate},}\ }\href@noop {} {\bibfield  {journal} {\bibinfo
   {journal} {Science}\ }\textbf {\bibinfo {volume} {327}},\ \bibinfo {pages}
  {155--156} (\bibinfo {year} {2010})}\BibitemShut {NoStop}%
\bibitem [{\citenamefont {Miyakawa}\ \emph
  {et~al.}(2008{\natexlab{a}})\citenamefont {Miyakawa}, \citenamefont {Sogo},\
  and\ \citenamefont {Pu}}]{Miyakawa:2008}%
  \BibitemOpen
  \bibfield  {author} {\bibinfo {author} {\bibfnamefont {T.}~\bibnamefont
  {Miyakawa}}, \bibinfo {author} {\bibfnamefont {T.}~\bibnamefont {Sogo}}, \
  and\ \bibinfo {author} {\bibfnamefont {H.}~\bibnamefont {Pu}},\ }\bibfield
  {title} {\enquote {\bibinfo {title} {Phase-space deformation of a trapped
  dipolar {Fermi} gas},}\ }\href@noop {} {\bibfield  {journal} {\bibinfo
  {journal} {Phys. Rev. A}\ }\textbf {\bibinfo {volume} {77}},\ \bibinfo
  {pages} {061603} (\bibinfo {year} {2008}{\natexlab{a}})}\BibitemShut
  {NoStop}%
\bibitem [{\citenamefont {Fregoso}\ \emph {et~al.}(2009)\citenamefont
  {Fregoso}, \citenamefont {Sun}, \citenamefont {Fradkin},\ and\ \citenamefont
  {Lev}}]{Fregoso:2009}%
  \BibitemOpen
  \bibfield  {author} {\bibinfo {author} {\bibfnamefont {B.~M.}\ \bibnamefont
  {Fregoso}}, \bibinfo {author} {\bibfnamefont {K.}~\bibnamefont {Sun}},
  \bibinfo {author} {\bibfnamefont {E.}~\bibnamefont {Fradkin}}, \ and\
  \bibinfo {author} {\bibfnamefont {B.~L.}\ \bibnamefont {Lev}},\ }\bibfield
  {title} {\enquote {\bibinfo {title} {Biaxial nematic phases in ultracold
  dipolar {Fermi} gases},}\ }\href@noop {} {\bibfield  {journal} {\bibinfo
  {journal} {New J. Phys.}\ }\textbf {\bibinfo {volume} {11}},\ \bibinfo
  {pages} {103003} (\bibinfo {year} {2009})}\BibitemShut {NoStop}%
\bibitem [{\citenamefont {Quintanilla}\ \emph {et~al.}(2009)\citenamefont
  {Quintanilla}, \citenamefont {Carr},\ and\ \citenamefont
  {Betouras}}]{Quintanilla:2009}%
  \BibitemOpen
  \bibfield  {author} {\bibinfo {author} {\bibfnamefont {J.}~\bibnamefont
  {Quintanilla}}, \bibinfo {author} {\bibfnamefont {S.~T.}\ \bibnamefont
  {Carr}}, \ and\ \bibinfo {author} {\bibfnamefont {J.~J.}\ \bibnamefont
  {Betouras}},\ }\bibfield  {title} {\enquote {\bibinfo {title} {Metanematic,
  smectic, and crystalline phases of dipolar fermions in an optical lattice},}\
  }\href@noop {} {\bibfield  {journal} {\bibinfo  {journal} {Phys. Rev. A}\
  }\textbf {\bibinfo {volume} {79}},\ \bibinfo {pages} {031601} (\bibinfo
  {year} {2009})}\BibitemShut {NoStop}%
\bibitem [{\citenamefont {Lin}\ \emph {et~al.}(2010)\citenamefont {Lin},
  \citenamefont {Zhao},\ and\ \citenamefont {Liu}}]{Liu10}%
  \BibitemOpen
  \bibfield  {author} {\bibinfo {author} {\bibfnamefont {C.}~\bibnamefont
  {Lin}}, \bibinfo {author} {\bibfnamefont {E.}~\bibnamefont {Zhao}}, \ and\
  \bibinfo {author} {\bibfnamefont {W.~V.}\ \bibnamefont {Liu}},\ }\bibfield
  {title} {\enquote {\bibinfo {title} {Liquid crystal phases of ultracold
  dipolar fermions on a lattice},}\ }\href {\doibase
  10.1103/PhysRevB.81.045115} {\bibfield  {journal} {\bibinfo  {journal} {Phys.
  Rev. B}\ }\textbf {\bibinfo {volume} {81}},\ \bibinfo {pages} {045115}
  (\bibinfo {year} {2010})}\BibitemShut {NoStop}%
\bibitem [{\citenamefont {Sun}\ \emph {et~al.}(2010)\citenamefont {Sun},
  \citenamefont {Wu},\ and\ \citenamefont {Das~Sarma}}]{DasSarma10c}%
  \BibitemOpen
  \bibfield  {author} {\bibinfo {author} {\bibfnamefont {K.}~\bibnamefont
  {Sun}}, \bibinfo {author} {\bibfnamefont {C.}~\bibnamefont {Wu}}, \ and\
  \bibinfo {author} {\bibfnamefont {S.}~\bibnamefont {Das~Sarma}},\ }\bibfield
  {title} {\enquote {\bibinfo {title} {Spontaneous inhomogeneous phases in
  ultracold dipolar {Fermi} gases},}\ }\href {\doibase
  10.1103/PhysRevB.82.075105} {\bibfield  {journal} {\bibinfo  {journal} {Phys.
  Rev. B}\ }\textbf {\bibinfo {volume} {82}},\ \bibinfo {pages} {075105}
  (\bibinfo {year} {2010})}\BibitemShut {NoStop}%
\bibitem [{\citenamefont {Fregoso}\ and\ \citenamefont
  {Fradkin}(2009)}]{Fregoso2009b}%
  \BibitemOpen
  \bibfield  {author} {\bibinfo {author} {\bibfnamefont {B.~M.}\ \bibnamefont
  {Fregoso}}\ and\ \bibinfo {author} {\bibfnamefont {E.}~\bibnamefont
  {Fradkin}},\ }\bibfield  {title} {\enquote {\bibinfo {title} {Ferronematic
  ground state of the dilute dipolar {Fermi} gas},}\ }\href@noop {} {\bibfield
  {journal} {\bibinfo  {journal} {Phys. Rev. Lett.}\ }\textbf {\bibinfo
  {volume} {103}},\ \bibinfo {pages} {205301} (\bibinfo {year}
  {2009})}\BibitemShut {NoStop}%
\bibitem [{\citenamefont {Fregoso}\ and\ \citenamefont
  {Fradkin}(2010)}]{fregoso:2010}%
  \BibitemOpen
  \bibfield  {author} {\bibinfo {author} {\bibfnamefont {B.~M.}\ \bibnamefont
  {Fregoso}}\ and\ \bibinfo {author} {\bibfnamefont {E.}~\bibnamefont
  {Fradkin}},\ }\bibfield  {title} {\enquote {\bibinfo {title} {Unconventional
  magnetism in imbalanced {Fermi} systems with magnetic dipolar
  interactions},}\ }\href@noop {} {\bibfield  {journal} {\bibinfo  {journal}
  {Phys. Rev. B}\ }\textbf {\bibinfo {volume} {81}},\ \bibinfo {pages} {214443}
  (\bibinfo {year} {2010})}\BibitemShut {NoStop}%
\bibitem [{\citenamefont {He}\ and\ \citenamefont
  {Hofstetter}(2011)}]{Hofstetter11}%
  \BibitemOpen
  \bibfield  {author} {\bibinfo {author} {\bibfnamefont {L.}~\bibnamefont
  {He}}\ and\ \bibinfo {author} {\bibfnamefont {W.}~\bibnamefont
  {Hofstetter}},\ }\bibfield  {title} {\enquote {\bibinfo {title} {Supersolid
  phase of cold fermionic polar molecules in two-dimensional optical
  lattices},}\ }\href {\doibase 10.1103/PhysRevA.83.053629} {\bibfield
  {journal} {\bibinfo  {journal} {Phys. Rev. A}\ }\textbf {\bibinfo {volume}
  {83}},\ \bibinfo {pages} {053629} (\bibinfo {year} {2011})}\BibitemShut
  {NoStop}%
\bibitem [{\citenamefont {{Bhongale}}\ \emph {et~al.}(2011)\citenamefont
  {{Bhongale}}, \citenamefont {{Mathey}}, \citenamefont {{Tsai}}, \citenamefont
  {{Clark}},\ and\ \citenamefont {{Zhao}}}]{Clark11}%
  \BibitemOpen
  \bibfield  {author} {\bibinfo {author} {\bibfnamefont {S.~G.}\ \bibnamefont
  {{Bhongale}}}, \bibinfo {author} {\bibfnamefont {L.}~\bibnamefont
  {{Mathey}}}, \bibinfo {author} {\bibfnamefont {S.-W.}\ \bibnamefont
  {{Tsai}}}, \bibinfo {author} {\bibfnamefont {C.~W.}\ \bibnamefont {{Clark}}},
  \ and\ \bibinfo {author} {\bibfnamefont {E.}~\bibnamefont {{Zhao}}},\
  }\bibfield  {title} {\enquote {\bibinfo {title} {{Bond order solid of
  two-dimensional dipolar fermions}},}\ }\href@noop {} {\  (\bibinfo {year}
  {2011})},\ \Eprint {http://arxiv.org/abs/1111.2873} {arXiv:1111.2873}
  \BibitemShut {NoStop}%
\bibitem [{\citenamefont {Chicireanu}\ \emph {et~al.}(2006)\citenamefont
  {Chicireanu}, \citenamefont {Pouderous}, \citenamefont {Barb\'e},
  \citenamefont {Laburthe-Tolra}, \citenamefont {Mar\'echal}, \citenamefont
  {Vernac}, \citenamefont {Keller},\ and\ \citenamefont {Gorceix}}]{Gorceix06}%
  \BibitemOpen
  \bibfield  {author} {\bibinfo {author} {\bibfnamefont {R.}~\bibnamefont
  {Chicireanu}}, \bibinfo {author} {\bibfnamefont {A.}~\bibnamefont
  {Pouderous}}, \bibinfo {author} {\bibfnamefont {R.}~\bibnamefont {Barb\'e}},
  \bibinfo {author} {\bibfnamefont {B.}~\bibnamefont {Laburthe-Tolra}},
  \bibinfo {author} {\bibfnamefont {E.}~\bibnamefont {Mar\'echal}}, \bibinfo
  {author} {\bibfnamefont {L.}~\bibnamefont {Vernac}}, \bibinfo {author}
  {\bibfnamefont {J.-C.}\ \bibnamefont {Keller}}, \ and\ \bibinfo {author}
  {\bibfnamefont {O.}~\bibnamefont {Gorceix}},\ }\bibfield  {title} {\enquote
  {\bibinfo {title} {Simultaneous magneto-optical trapping of bosonic and
  fermionic chromium atoms},}\ }\href {\doibase 10.1103/PhysRevA.73.053406}
  {\bibfield  {journal} {\bibinfo  {journal} {Phys. Rev. A}\ }\textbf {\bibinfo
  {volume} {73}},\ \bibinfo {pages} {053406} (\bibinfo {year}
  {2006})}\BibitemShut {NoStop}%
\bibitem [{\citenamefont {Berglund}\ \emph {et~al.}(2007)\citenamefont
  {Berglund}, \citenamefont {Lee},\ and\ \citenamefont
  {McClelland}}]{Berglund:2007}%
  \BibitemOpen
  \bibfield  {author} {\bibinfo {author} {\bibfnamefont {A.~J.}\ \bibnamefont
  {Berglund}}, \bibinfo {author} {\bibfnamefont {S.~A.}\ \bibnamefont {Lee}}, \
  and\ \bibinfo {author} {\bibfnamefont {J.~J.}\ \bibnamefont {McClelland}},\
  }\bibfield  {title} {\enquote {\bibinfo {title} {Sub-{Doppler} laser cooling
  and magnetic trapping of erbium},}\ }\href@noop {} {\bibfield  {journal}
  {\bibinfo  {journal} {Phys. Rev. A}\ }\textbf {\bibinfo {volume} {76}},\
  \bibinfo {pages} {053418} (\bibinfo {year} {2007})}\BibitemShut {NoStop}%
\bibitem [{\citenamefont {Ni}\ \emph {et~al.}(2010)\citenamefont {Ni},
  \citenamefont {Ospelkaus}, \citenamefont {Wang}, \citenamefont {Quemener},
  \citenamefont {Neyenhuis}, \citenamefont {{de Miranda}}, \citenamefont
  {Bohn}, \citenamefont {Ye},\ and\ \citenamefont {Jin}}]{Ni2010}%
  \BibitemOpen
  \bibfield  {author} {\bibinfo {author} {\bibfnamefont {K.~K.}\ \bibnamefont
  {Ni}}, \bibinfo {author} {\bibfnamefont {S.}~\bibnamefont {Ospelkaus}},
  \bibinfo {author} {\bibfnamefont {D.}~\bibnamefont {Wang}}, \bibinfo {author}
  {\bibfnamefont {G.}~\bibnamefont {Quemener}}, \bibinfo {author}
  {\bibfnamefont {B.}~\bibnamefont {Neyenhuis}}, \bibinfo {author}
  {\bibfnamefont {M.~H.~G.}\ \bibnamefont {{de Miranda}}}, \bibinfo {author}
  {\bibfnamefont {J.~L.}\ \bibnamefont {Bohn}}, \bibinfo {author}
  {\bibfnamefont {J.}~\bibnamefont {Ye}}, \ and\ \bibinfo {author}
  {\bibfnamefont {D.~S.}\ \bibnamefont {Jin}},\ }\bibfield  {title} {\enquote
  {\bibinfo {title} {Dipolar collisions of polar molecules in the quantum
  regime},}\ }\href@noop {} {\bibfield  {journal} {\bibinfo  {journal}
  {Nature}\ }\textbf {\bibinfo {volume} {464}},\ \bibinfo {pages} {1324--1328}
  (\bibinfo {year} {2010})}\BibitemShut {NoStop}%
\bibitem [{\citenamefont {{Chotia}}\ \emph {et~al.}(2011)\citenamefont
  {{Chotia}}, \citenamefont {{Neyenhuis}}, \citenamefont {{Moses}},
  \citenamefont {{Yan}}, \citenamefont {{Covey}}, \citenamefont {{Foss-Feig}},
  \citenamefont {{Rey}}, \citenamefont {{Jin}},\ and\ \citenamefont
  {{Ye}}}]{Chotia11}%
  \BibitemOpen
  \bibfield  {author} {\bibinfo {author} {\bibfnamefont {A.}~\bibnamefont
  {{Chotia}}}, \bibinfo {author} {\bibfnamefont {B.}~\bibnamefont
  {{Neyenhuis}}}, \bibinfo {author} {\bibfnamefont {S.~A.}\ \bibnamefont
  {{Moses}}}, \bibinfo {author} {\bibfnamefont {B.}~\bibnamefont {{Yan}}},
  \bibinfo {author} {\bibfnamefont {J.~P.}\ \bibnamefont {{Covey}}}, \bibinfo
  {author} {\bibfnamefont {M.}~\bibnamefont {{Foss-Feig}}}, \bibinfo {author}
  {\bibfnamefont {A.~M.}\ \bibnamefont {{Rey}}}, \bibinfo {author}
  {\bibfnamefont {D.~S.}\ \bibnamefont {{Jin}}}, \ and\ \bibinfo {author}
  {\bibfnamefont {J.}~\bibnamefont {{Ye}}},\ }\bibfield  {title} {\enquote
  {\bibinfo {title} {{Long-lived dipolar molecules and Feshbach molecules in a
  3D optical lattice}},}\ }\href@noop {} {\  (\bibinfo {year} {2011})},\
  \Eprint {http://arxiv.org/abs/1110.4420} {arXiv:1110.4420} \BibitemShut
  {NoStop}%
\bibitem [{\citenamefont {Lu}\ \emph {et~al.}(2011)\citenamefont {Lu},
  \citenamefont {Burdick}, \citenamefont {Youn},\ and\ \citenamefont
  {Lev}}]{Lu2011b}%
  \BibitemOpen
  \bibfield  {author} {\bibinfo {author} {\bibfnamefont {M.}~\bibnamefont
  {Lu}}, \bibinfo {author} {\bibfnamefont {N.~Q.}\ \bibnamefont {Burdick}},
  \bibinfo {author} {\bibfnamefont {S.-H.}\ \bibnamefont {Youn}}, \ and\
  \bibinfo {author} {\bibfnamefont {B.~L.}\ \bibnamefont {Lev}},\ }\bibfield
  {title} {\enquote {\bibinfo {title} {Strongly dipolar {Bose-Einstein}
  condensate of dysprosium},}\ }\href {\doibase 10.1103/PhysRevLett.107.190401}
  {\bibfield  {journal} {\bibinfo  {journal} {Phys. Rev. Lett.}\ }\textbf
  {\bibinfo {volume} {107}},\ \bibinfo {pages} {190401} (\bibinfo {year}
  {2011})}\BibitemShut {NoStop}%
\bibitem [{\citenamefont {Pasquiou}\ \emph {et~al.}(2010)\citenamefont
  {Pasquiou}, \citenamefont {Bismut}, \citenamefont {Beaufils}, \citenamefont
  {Crubellier}, \citenamefont {Mar\'echal}, \citenamefont {Pedri},
  \citenamefont {Vernac}, \citenamefont {Gorceix},\ and\ \citenamefont
  {Laburthe-Tolra}}]{Pasquiou10}%
  \BibitemOpen
  \bibfield  {author} {\bibinfo {author} {\bibfnamefont {B.}~\bibnamefont
  {Pasquiou}}, \bibinfo {author} {\bibfnamefont {G.}~\bibnamefont {Bismut}},
  \bibinfo {author} {\bibfnamefont {Q.}~\bibnamefont {Beaufils}}, \bibinfo
  {author} {\bibfnamefont {A.}~\bibnamefont {Crubellier}}, \bibinfo {author}
  {\bibfnamefont {E.}~\bibnamefont {Mar\'echal}}, \bibinfo {author}
  {\bibfnamefont {P.}~\bibnamefont {Pedri}}, \bibinfo {author} {\bibfnamefont
  {L.}~\bibnamefont {Vernac}}, \bibinfo {author} {\bibfnamefont
  {O.}~\bibnamefont {Gorceix}}, \ and\ \bibinfo {author} {\bibfnamefont
  {B.}~\bibnamefont {Laburthe-Tolra}},\ }\bibfield  {title} {\enquote {\bibinfo
  {title} {Control of dipolar relaxation in external fields},}\ }\href
  {\doibase 10.1103/PhysRevA.81.042716} {\bibfield  {journal} {\bibinfo
  {journal} {Phys. Rev. A}\ }\textbf {\bibinfo {volume} {81}},\ \bibinfo
  {pages} {042716} (\bibinfo {year} {2010})}\BibitemShut {NoStop}%
\bibitem [{\citenamefont {Giovanazzi}\ \emph {et~al.}(2002)\citenamefont
  {Giovanazzi}, \citenamefont {{G\"{o}rlitz}},\ and\ \citenamefont
  {Pfau}}]{Pfau02}%
  \BibitemOpen
  \bibfield  {author} {\bibinfo {author} {\bibfnamefont {S.}~\bibnamefont
  {Giovanazzi}}, \bibinfo {author} {\bibfnamefont {A.}~\bibnamefont
  {{G\"{o}rlitz}}}, \ and\ \bibinfo {author} {\bibfnamefont {T.}~\bibnamefont
  {Pfau}},\ }\bibfield  {title} {\enquote {\bibinfo {title} {Tuning the dipolar
  interaction in quantum gases},}\ }\href@noop {} {\bibfield  {journal}
  {\bibinfo  {journal} {Phys. Rev. Lett.}\ }\textbf {\bibinfo {volume}
  {\textbf{89}}},\ \bibinfo {pages} {130401} (\bibinfo {year}
  {2002})}\BibitemShut {NoStop}%
\bibitem [{\citenamefont {Dalmonte}\ \emph {et~al.}(2011)\citenamefont
  {Dalmonte}, \citenamefont {Zoller},\ and\ \citenamefont
  {Pupillo}}]{Pupillo11}%
  \BibitemOpen
  \bibfield  {author} {\bibinfo {author} {\bibfnamefont {M.}~\bibnamefont
  {Dalmonte}}, \bibinfo {author} {\bibfnamefont {P.}~\bibnamefont {Zoller}}, \
  and\ \bibinfo {author} {\bibfnamefont {G.}~\bibnamefont {Pupillo}},\
  }\bibfield  {title} {\enquote {\bibinfo {title} {Trimer liquids and crystals
  of polar molecules in coupled wires},}\ }\href {\doibase
  10.1103/PhysRevLett.107.163202} {\bibfield  {journal} {\bibinfo  {journal}
  {Phys. Rev. Lett.}\ }\textbf {\bibinfo {volume} {107}},\ \bibinfo {pages}
  {163202} (\bibinfo {year} {2011})}\BibitemShut {NoStop}%
\bibitem [{\citenamefont {Gorshkov}\ \emph {et~al.}(2011)\citenamefont
  {Gorshkov}, \citenamefont {Manmana}, \citenamefont {Chen}, \citenamefont
  {Ye}, \citenamefont {Demler}, \citenamefont {Lukin},\ and\ \citenamefont
  {Rey}}]{Gorshkov11}%
  \BibitemOpen
  \bibfield  {author} {\bibinfo {author} {\bibfnamefont {A.~V.}\ \bibnamefont
  {Gorshkov}}, \bibinfo {author} {\bibfnamefont {S.~R.}\ \bibnamefont
  {Manmana}}, \bibinfo {author} {\bibfnamefont {G.}~\bibnamefont {Chen}},
  \bibinfo {author} {\bibfnamefont {J.}~\bibnamefont {Ye}}, \bibinfo {author}
  {\bibfnamefont {E.}~\bibnamefont {Demler}}, \bibinfo {author} {\bibfnamefont
  {M.~D.}\ \bibnamefont {Lukin}}, \ and\ \bibinfo {author} {\bibfnamefont
  {A.~M.}\ \bibnamefont {Rey}},\ }\bibfield  {title} {\enquote {\bibinfo
  {title} {Tunable superfluidity and quantum magnetism with ultracold polar
  molecules},}\ }\href {\doibase 10.1103/PhysRevLett.107.115301} {\bibfield
  {journal} {\bibinfo  {journal} {Phys. Rev. Lett.}\ }\textbf {\bibinfo
  {volume} {107}},\ \bibinfo {pages} {115301} (\bibinfo {year}
  {2011})}\BibitemShut {NoStop}%
\bibitem [{\citenamefont {Kotochigova}\ and\ \citenamefont
  {Petrov}(2011)}]{Svetlana11}%
  \BibitemOpen
  \bibfield  {author} {\bibinfo {author} {\bibfnamefont {S.}~\bibnamefont
  {Kotochigova}}\ and\ \bibinfo {author} {\bibfnamefont {A.}~\bibnamefont
  {Petrov}},\ }\bibfield  {title} {\enquote {\bibinfo {title} {Anisotropy in
  the interaction of ultracold dysprosium},}\ }\href@noop {} {\bibfield
  {journal} {\bibinfo  {journal} {Phys. Chem. Chem. Phys.}\ }\textbf {\bibinfo
  {volume} {13}},\ \bibinfo {pages} {19165--19170} (\bibinfo {year}
  {2011})}\BibitemShut {NoStop}%
\bibitem [{\citenamefont {Lu}\ \emph {et~al.}(2010)\citenamefont {Lu},
  \citenamefont {Youn},\ and\ \citenamefont {Lev}}]{Lu2010}%
  \BibitemOpen
  \bibfield  {author} {\bibinfo {author} {\bibfnamefont {M.}~\bibnamefont
  {Lu}}, \bibinfo {author} {\bibfnamefont {S.-H.}\ \bibnamefont {Youn}}, \ and\
  \bibinfo {author} {\bibfnamefont {Benjamin~L.}\ \bibnamefont {Lev}},\
  }\bibfield  {title} {\enquote {\bibinfo {title} {Trapping ultracold
  dysprosium: {A} highly magnetic gas for dipolar physics},}\ }\href@noop {}
  {\bibfield  {journal} {\bibinfo  {journal} {Phys. Rev. Lett.}\ }\textbf
  {\bibinfo {volume} {104}},\ \bibinfo {pages} {063001} (\bibinfo {year}
  {2010})}\BibitemShut {NoStop}%
\bibitem [{\citenamefont {DeMarco}(2001)}]{DeMarcoThesis}%
  \BibitemOpen
  \bibfield  {author} {\bibinfo {author} {\bibfnamefont {B.}~\bibnamefont
  {DeMarco}},\ }\href@noop {} {\enquote {\bibinfo {title} {{Ph.D. thesis:
  Quantum Behavior of an Atomic Fermi Gas}},}\ } (\bibinfo {year} {2001}),\
  \Eprint {http://arxiv.org/abs/University of Colorado at Boulder} {University
  of Colorado at Boulder} \BibitemShut {NoStop}%
\bibitem [{\citenamefont {Hensler}\ \emph {et~al.}(2003)\citenamefont
  {Hensler}, \citenamefont {Werner}, \citenamefont {Griesmaier}, \citenamefont
  {Schmidt}, \citenamefont {{G\"{o}rlitz}}, \citenamefont {Pfau}, \citenamefont
  {Giovanazzi},\ and\ \citenamefont {Rzazewski}}]{Hensler:2003}%
  \BibitemOpen
  \bibfield  {author} {\bibinfo {author} {\bibfnamefont {S.}~\bibnamefont
  {Hensler}}, \bibinfo {author} {\bibfnamefont {J.}~\bibnamefont {Werner}},
  \bibinfo {author} {\bibfnamefont {A.}~\bibnamefont {Griesmaier}}, \bibinfo
  {author} {\bibfnamefont {P.O.}\ \bibnamefont {Schmidt}}, \bibinfo {author}
  {\bibfnamefont {A.}~\bibnamefont {{G\"{o}rlitz}}}, \bibinfo {author}
  {\bibfnamefont {T.}~\bibnamefont {Pfau}}, \bibinfo {author} {\bibfnamefont
  {S.}~\bibnamefont {Giovanazzi}}, \ and\ \bibinfo {author} {\bibfnamefont
  {K.}~\bibnamefont {Rzazewski}},\ }\bibfield  {title} {\enquote {\bibinfo
  {title} {Dipolar relaxation in an ultra-cold gas of magnetically trapped
  chromium atoms},}\ }\href@noop {} {\bibfield  {journal} {\bibinfo  {journal}
  {Appl. Phys. B}\ }\textbf {\bibinfo {volume} {77}},\ \bibinfo {pages}
  {765--772} (\bibinfo {year} {2003})}\BibitemShut {NoStop}%
\bibitem [{\citenamefont {Berglund}\ \emph {et~al.}(2008)\citenamefont
  {Berglund}, \citenamefont {Hanssen},\ and\ \citenamefont
  {McClelland}}]{Berglund:2008}%
  \BibitemOpen
  \bibfield  {author} {\bibinfo {author} {\bibfnamefont {A.~J.}\ \bibnamefont
  {Berglund}}, \bibinfo {author} {\bibfnamefont {J.~L.}\ \bibnamefont
  {Hanssen}}, \ and\ \bibinfo {author} {\bibfnamefont {J.~J.}\ \bibnamefont
  {McClelland}},\ }\bibfield  {title} {\enquote {\bibinfo {title} {Narrow-line
  magneto-optical cooling and trapping of strongly magnetic atoms},}\
  }\href@noop {} {\bibfield  {journal} {\bibinfo  {journal} {Phys. Rev. Lett.}\
  }\textbf {\bibinfo {volume} {100}},\ \bibinfo {pages} {113002} (\bibinfo
  {year} {2008})}\BibitemShut {NoStop}%
\bibitem [{\citenamefont {Miyakawa}\ \emph
  {et~al.}(2008{\natexlab{b}})\citenamefont {Miyakawa}, \citenamefont {Sogo},\
  and\ \citenamefont {Pu}}]{Pu08}%
  \BibitemOpen
  \bibfield  {author} {\bibinfo {author} {\bibfnamefont {T.}~\bibnamefont
  {Miyakawa}}, \bibinfo {author} {\bibfnamefont {T.}~\bibnamefont {Sogo}}, \
  and\ \bibinfo {author} {\bibfnamefont {H.}~\bibnamefont {Pu}},\ }\bibfield
  {title} {\enquote {\bibinfo {title} {Phase-space deformation of a trapped
  dipolar {Fermi} gas},}\ }\href {\doibase 10.1103/PhysRevA.77.061603}
  {\bibfield  {journal} {\bibinfo  {journal} {Phys. Rev. A}\ }\textbf {\bibinfo
  {volume} {77}},\ \bibinfo {pages} {061603} (\bibinfo {year}
  {2008}{\natexlab{b}})}\BibitemShut {NoStop}%
\bibitem [{\citenamefont {Bohn}\ \emph {et~al.}(2009)\citenamefont {Bohn},
  \citenamefont {Cavagnero},\ and\ \citenamefont {Ticknor}}]{Bohn09}%
  \BibitemOpen
  \bibfield  {author} {\bibinfo {author} {\bibfnamefont {J.~L.}\ \bibnamefont
  {Bohn}}, \bibinfo {author} {\bibfnamefont {M.}~\bibnamefont {Cavagnero}}, \
  and\ \bibinfo {author} {\bibfnamefont {C.}~\bibnamefont {Ticknor}},\
  }\bibfield  {title} {\enquote {\bibinfo {title} {Quasi-universal dipolar
  scattering in cold and ultracold gases},}\ }\href@noop {} {\bibfield
  {journal} {\bibinfo  {journal} {New J. Phys.}\ }\textbf {\bibinfo {volume}
  {11}},\ \bibinfo {pages} {055039} (\bibinfo {year} {2009})}\BibitemShut
  {NoStop}%
\end{thebibliography}
\end{document}